\documentclass[rnote]{aa}
\usepackage{graphicx}
\usepackage{multirow}
\usepackage{txfonts}
\usepackage{floatflt}
\usepackage[authoryear]{natbib}       
\bibpunct{(}{)}{;}{a}{}{,}
\begin{document}
\title{Strong linear polarization of V4332 Sgr:
a dusty disc geometry\thanks{Based on observations made with 
the Nordic Optical Telescope, operated on the island of La Palma jointly by Denmark, 
Finland, Iceland, Norway, and Sweden, in the Spanish Observatorio del Roque de 
los Muchachos of the Instituto de Astrofisica de Canarias.}}
\author{T. Kami\'nski\inst{1,2} \and R. Tylenda\inst{1}}  

\offprints{T. Kami\'{n}ski}
\institute{Department for Astrophysics, N. Copernicus
            Astronomical Center, Rabia\'{n}ska 8,
            87-100 Toru\'{n}, Poland\\ 
            \email{tomkam, tylenda@ncac.torun.pl}
       \and currently at Max-Planck Institut f\"ur Radioastronomie, Auf dem
            H\"ugel 69, 53121 Bonn, Germany}
\date{Received; accepted}
\abstract
{The eruption of V4332 Sgr was observed in 1994. During the
outburst, the object became extremely red, so it is considered as
belonging to luminous red transients of the V838~Mon type.
It has recently been suggested that the central object in V4332~Sgr is now
 hidden in a dusty disc and that the photospheric spectrum of this object
observed in the optical results from scattering the central star radiation
on dust grains in the disc.} 
{One expects significant polarization of the
spectrum in this case. We investigate this prediction.} 
{We present and analyse polarimetric observations of V4332~Sgr in the $V$
 and $R$ photometric bands done with the NOT telescope.}
{The optical light of V4332 Sgr is linearly polarized with a degree of
$\sim$26\% in the $V$ band and $\sim$11\% in $R$.}
{Discussion of the observed polarization leads us to conclude that the
photospheric spectrum observed in V4332~Sgr is probably produced by dust scattering
not only in the disc but also in the outflow from the object seen in the
emission features.}

        
\titlerunning{Polarization of V4332 Sgr}
\authorrunning{Kami\'nski \& Tylenda}
\maketitle
\section{Introduction}
 The eruption of V4332 Sgr was observed in 1994 \citep{martini}.
Discovered as a possible nova, the object appeared to be unusual,
mainly because of its spectral evolution. 
It quickly evolved from K to M spectral types and after a month it declined as very late
M-type giant. The object is now classified as a red optical transient of
the V838~Mon type \citep{muna,tcgs}.
As discussed by \citet{ts06},
thermonuclear mechanisms (classical nova, He-shell flash)
probably cannot explain the
observed outbursts of these objects. The stellar
collision-merger scenario proposed in \citet{st03} and
further developed in \citet{ts06} is a promising
hypothesis for explaining the nature of these eruptions. 
 
Now more than a decade after outburst, V4332~Sgr continues
to present unusual observational characteristics. Apart from displaying an M-type
continuum in the optical,
it shows a unique emission line spectrum of very low
excitation \citep{tcgs,kimes}. The object is also very bright in the
infrared \citep{tcgs,baner07}.

In a recent paper, after having analysed the optical emission line
spectrum and the spectral energy distribution of V4332~Sgr, \citet{kst}
conclude that the main object, probably an M-type giant, is hidden in a
circumstellar disc seen almost edge-on. The stellar-like spectrum observed
in the optical probably results from scattering the central star light on
dust grains at the outer edge of the disc. This implies that the optical
spectrum should display a significant polarization. In the present paper we
report on polarimetric measurements done in the $V$ and $R$
photometric bands, which show that the optical light from V4332~Sgr is
indeed strongly polarized.

\section{Observations and data processing}

Polarimetric observations of V4332 Sgr were obtained on August 8, 2010 at the
2.56~m Nordic Optical Telescope using the Andalucia Faint Object Spectrograph 
and Camera (ALFOSC). 
Polarimetry was performed in the Johnson $V$ and Kron-Cousins $R$ bands by inserting 
a half-wave plate and a calcite plate in the light beam. A calcite plate splits 
the incident light beam into two beams of orthogonal polarization, i.e. ordinary ($o$) 
and extraordinary ($e$) beams. For ALFOSC, this provides two overlapping images of 
the field of view  separated by 15 arcsec. For each filter, a sequence of 8 exposures 
was obtained with the half-wave plate rotated by 22\fdg5 between 0\degr\ and 157\fdg5. 
The exposure times per image were 400~s in the $V$ band and 70~s in $R$. 

Data were reduced using IRAF\footnote{IRAF is distributed by the National Optical 
Astronomy Observatories,
which are operated by the Association of Universities for Research
in Astronomy, Inc., under cooperative agreement with the National
Science Foundation}. All the CCD frames were corrected for bias, flat-field, and 
cosmic-ray effects.
The object's signal was measured for the ordinary ($f^o$) and 
extraordinary ($f^e$) beam on each image . The signal was integrated within circular apertures 
(of the same size for each exposure), and sky background
was subtracted. 

For each retarder angle, $\theta_i$, the normalized flux difference was
calculated,
\begin{equation}
f(\theta_i)=\frac{f^o(\theta_i)-f^e(\theta_i)}{f^o(\theta_i)+f^e(\theta_i)},
\end{equation}
from which reduced (normalized) Stokes parameters $P_Q$ and $P_U$ were derived 
\citep[see][]{fossati} 
\begin{eqnarray}
P_Q(0,45\degr)&=&\frac{f(0\degr)-f(45\degr)}{2},\\
P_Q(90,135\degr)&=&\frac{f(90\degr)-f(135\degr)}{2},\\
P_U(22\fdg5,67\fdg5)&=&\frac{f(22\fdg5)-f(67\fdg5)}{2},\\
P_U(112\fdg5,157\fdg5)&=&\frac{f(112\fdg5)-f(157\fdg5)}{2}.
\end{eqnarray}
The reduced parameters are in the following relation to the standard $IQU$ Stokes 
parameters
\begin{equation}
P_Q=\frac{Q}{I},\\
 P_U=\frac{U}{I}.
\end{equation}
For the 8-angle sequence of retarder positions, two values for each of the reduced 
Stokes parameters are obtained. For each filter we took average values of $P_Q$ 
and $P_U$ to derive the degree of linear polarization, $P_L$, and the position 
angle $PA$ of the maximum-polarization vector \citep[see][]{landi}
\begin{equation}
P_L=\sqrt{P_Q^2+P_U^2}\cdot 100\%,\\
PA=\frac{1}{2} {\rm sign}(P_U) \arccos \biggl(\frac{P_Q}{P_L}\biggr).
\end{equation}

Uncertainties were first estimated as standard deviations of the background counts 
propagated in the calculation of $P_L$ and $PA$ \citep[see][]{fossati}. 
Those errors, however, turned out to be insignificant when compared to errors related 
to the deviations between the two sets of the derived Stokes parameters $P_Q$ and $P_U$. 
Below we provide those deviations (in the sense of 1$\sigma$) propagated to the 
calculation of $P_L$ and $PA$ as our final uncertainties.

The instrumental polarization was checked by observations of a high polarization 
standard, BD+64\,106, and a low polarization standard, BD+32\,3739. 
For the latter star we found $P_L(V)$=(0.08$\pm$0.07)\% and $P_L(R)$=(0.04$\pm$0.09)\% 
(for the $V$ and $R$ bands, respectively)\footnote{After correcting these results for 
bias, which is related to $P_L$ being a positive definite quantity, 
and following instructions in \citet{simmons}, we got corresponding 1$\sigma$ 
confidence ranges of $P_L(V)$=0.0$\div$0.13\% and $P_L(R)$=0.0$\div$0.07\%.}, 
which is consistent with the catalogue polarization of this object \citep{zeropol}. 
For the high-polarization standard we found  $P_L(V)$=(5.50$\pm$0.18)\% and 
$P_L(R)$=(5.30$\pm$0.01)\%, which can be compared with the catalogue values
of $P_L(V)$=(5.69$\pm$0.04)\% and $P_L(R)$=(5.15$\pm$0.10)\% \citep{highpol}.  
We also measured  polarization of a few field stars in the frames, 
and they all showed no or a marginal polarization within measurement 
uncertainties. We can therefore conclude that there is no significant instrument 
polarization that could affect the derived values of $P_L$.

The estimated position angles for the high-polarization standard are 
$PA(V)$=101\fdg2$\pm$0\fdg5 and $PA(R)$=97\fdg4$\pm$0\fdg5, which can be compared 
to the catalogue values of $PA_0(V)$=96\fdg6$\pm$0\fdg2 and 
$PA_0(R)$=96\fdg7$\pm$0\fdg5 \citep{highpol}. The differences between 
the derived and standard values 
can be interpreted as due to instrumental polarization. We treat 
the differences $ZP$=$PA_0-PA$ as zero-points for the scale of PA, 
i.e. $ZP(V)$=--4\fdg58 and $ZP(R)$=--0\fdg68.  

\section{Results}

V4332 Sgr is at present very faint in the optical 
($V\approx19.8$~mag, $R\approx17.9$~mag). Nevertheless we have found that
the polarization of V4332 Sgr is significant in both bands. We
obtained $P_L(V)$=(25.8$\pm$3.0)\% and $P_L(R)$=(11.3$\pm$2.0)\%. 
The derived position angles are $PA(V)$=113\fdg5$\pm$3\fdg3 and 
$PA(R)$=116\fdg2$\pm$4\fdg8, which  
gives $PA(V)$=108\fdg9$\pm$4\fdg0 and $PA(R)$=115\fdg9$\pm$5\fdg4 
when corrected for the instrument
polarization (see above).
(Here errors also include contributions from the uncertainties of 
the derived $PA$ zero points.) 
A weighted mean of 
the two angles gives $PA$=111\fdg3$\pm$3\fdg2.

\section{Discussion}

\begin{figure}
\includegraphics[angle=270,width=\columnwidth]{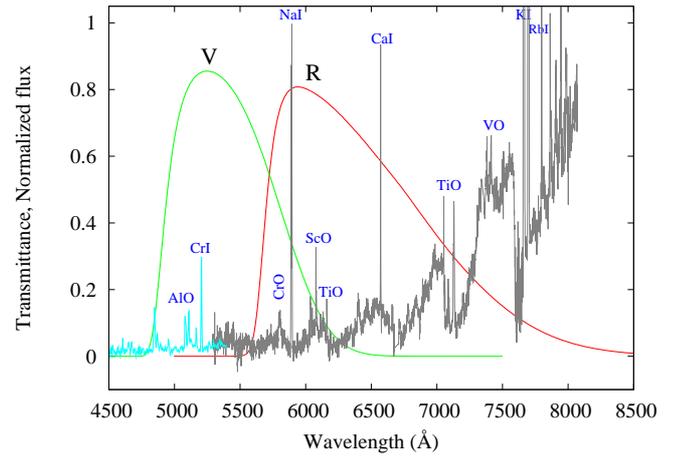}
\caption{The thick lines (green and red) are transmission curves of the
$V$ and  $R$ 
filters used in the polarimetric observations. Spectra of V4332 Sgr are 
shown for comparison. The part for $\lambda<5400$~\AA\ is a low-resolution spectrum 
obtained in October 2005 in SAAO (unpublished).
The spectrum covering the remaining wavelength range was obtained with 
the Subaru Telescope in June 2009 and is described in \citet{kst}. 
Spectra are normalized to the peaks of the \ion{Na}{I} doublet at 5890~\AA. 
}
\label{fig}
\end{figure}

Our polarimetric observations 
were performed in the spectral regions, where, apart from the stellar
(photospheric) spectrum, emission features
significantly contribute to the photometric fluxes, as
can be seen from Fig.~\ref{fig}. 
We estimated that the atomic and molecular emission lines constitute 
$\sim$40\% of the total fluxes in both bands.
 As discussed in \citet{kst}, the emission features are
produced in the radiation pumping mechanism  most probably in 
a matter directly visible by the observer.
Therefore no polarization in 
the lines is expected. If this is the case, the polarization 
degree of the continuum would be $\sim$40\% and $\sim$20\% in the $V$ and
$R$ bands, respectively.
In either case, spectropolarimetric observations would be of particular
importance for a more conclusive discussion of the origin of the
polarization and the nature of V4332~Sgr.

Regardless of whether the polarization is limited to the continuum or not, 
the observed degree of polarization of V4332 Sgr is very high. This limits 
the choice of polarization mechanisms. It is difficult to explain that high polarization 
with dichroic extinction caused by aligned non-spherical grains. There is 
no known alignment mechanism, which could give the required high alignment efficiency 
\citep[e.g.][]{bastien}. The only alternative is the light scattering in 
circumstellar material. This is consistent with the work of \citet{kst}, where
the authors postulate that the central M-type giant in V4332~Sgr is hidden
in a dusty disc seen almost edge-on. The stellar-like spectrum observed in
the optical would in fact be the giant spectrum scattered on dust grains in the disc.
In this case, forward scattering 
on big dust grains at the outer edge of the disc is most effective in producing 
the observed radiation.
However, the expected polarization degree is small, if scattering occurs
at small angels. 
Radiative transfer modelling of scattering in a disc seen at inclination angles 
80\degr$\leq i \leq$ 90\degr\ done in \citet{bastien} predicts 
a maximum polarization degree of 8--11\%.
Gomez's Hamburger (IRAS 18059-3211) is a good example of a high-inclination disc 
scattering light from the central star \citep{wood08}. This object shows a maximum 
polarization of 20--30\% in outer faint regions of the observed image
\citep{ruiz}, i.e., regions produced by scattering at large angles.
In the central, bright regions, produced by forward scattering, the observed
polarization is much lower. If the image of Gomez's Hamburger were not
resolved, as for V4332~Sgr, the net polarization would have
been observed at $\sim$10\% or less.

\citet{kst} also postulate that there is an outflowing matter in V4332~Sgr,
where the observed emission features of atoms and molecules are formed by
absorption and re-emission of the radiation from the central object.
This matter is cold, as indicated by the profiles of the emission molecular
bands analysed in \citet{kst}, and most probably contains dust. Thus the observed
optical continuum can also be partly produced in the outflow scattering the
central star radiation. If the outflow is concentrated along the disc axis, 
scattering would occur at angles close to
90\degr, leading to a considerable polarization of the observed spectrum.
\citet{bastien} \citep[see also][]{solc} 
found that for bipolar outflows from discs with inclination angles of 
80--90\degr\
the maximum polarization can reach 60--94\%. Therefore the high
polarization of the observed spectrum of V4332~Sgr may indicate that this
spectrum is produced by dust scattering not only in the disc but also
in the matter outflowing along the disc axis.

 In either case, the present polarization measurements confirm the conclusion of
\citet{kst} that the main object in V4332~Sgr is now hidden in a dusty disc
seen almost edge-on. This implies that the image of the object should
display a dark lane across the image, as it observed e.g. in
Gomez's Humburger. In November~1997 an image of V4332~Sgr was taken by HST
\citep[briefly discussed in][]{kimes}. No dark lane is seen but the image
profile is not significantly different from that of the point source. With
the lower limit distance of 1~kpc, as derived in \citet{kst}, the angular 
resolution of HST of 0.1~arcsec implies
that the linear dimensions of the V4332~Sgr disc are $\la$100~AU. We do not
know the nature of the disc, nor when it was formed. If it resulted from the
1994 eruption, e.g. as a remnant of a binary merger, its dimensions are
expected to be much smaller that the above upper limit.

\acknowledgements{The data presented here have been taken using ALFOSC, 
which is owned by the Instituto de Astrofisica de Andalucia (IAA) and 
operated at the Nordic Optical Telescope under agreement between IAA and 
the NBIfAFG of the Astronomical Observatory of Copenhagen. The research
reported on in this paper has partly been supported by a grant no.
N\,N203\,403939 financed by the Polish Ministry of Sciences and Higher
Education. }
   
\bibliographystyle{aa}

\end{document}